\documentclass{iopart}
\usepackage{iopams}
\usepackage{amssymb}
\expandafter\let\csname equation*\endcsname\relax
\expandafter\let\csname endequation*\endcsname\relax
\usepackage{amsmath}
\usepackage{bm}
\usepackage{graphicx}
\begin{document}

\title{Heisenberg uncertainty relation for relativistic electrons}
\author{Iwo Bialynicki-Birula}
\address{Center for Theoretical Physics, Polish Academy of Sciences\\
Aleja Lotnik\'ow 32/46, 02-668 Warsaw, Poland\ead{birula@cft.edu.pl}}
\author{Zofia Bialynicka-Birula}
\address{Institute of Physics, Polish Academy of Sciences\\
Aleja Lotnik\'ow 32/46, 02-668 Warsaw, Poland}

\begin{abstract}
The Heisenberg uncertainty relation is derived for relativistic electrons described by the Dirac equation. The standard nonrelativistic lower bound $3/2\hbar$ is attained only in the limit and the wave function that reproduces this value is singular. At the other end, in the ultrarelativistic limit, the bound is the same as that found before for photons.
\end{abstract}
\submitto{New Journal of Physics}

\section{Introduction}
Recent progress in the generation of relativistic beams of electrons calls for adequate theoretical tools, based on the Dirac equation, that would allow to account precisely for quantum properties of these electrons. The problem raised and answered in this paper is that of Heisenberg uncertainty relation (HUR) for relativistic electrons. It is shown that the relativistic HUR is markedly different from its original nonrelativistic form.

In most monographs \cite{bs}-\cite{th} devoted to relativistic quantum mechanics of electrons the Heisenberg uncertainty relations has not been mentioned. In those few places where there is a reference to uncertainty relations in a relativistic theory, the authors \cite{wp,blp} follow the early paper by Landau and Peierls \cite{lp} giving only a qualitative analysis of limitations imposed by relativity on measurements. In this work we present a complementary approach to uncertainty relations. It could be said that we do what Kennard \cite{ken} had done for the nonrelativistic form of the uncertainty relation. He derived an exact mathematical inequality which expressed precisely the heuristic explanation (Bohr-Heisenberg microscope) of the nonrelativistic uncertainty relation. We do the same for relativistic electrons.

In nonrelativistic quantum mechanics one may rely directly on the interpretation of the modulus squared of the wave function $|\psi(\bm r,t)|^2$ and its Fourier transform $|{\tilde\psi(\bm p,t)}|^2$ as the probability densities in position and momentum space. In relativistic quantum mechanics there is an ambiguity in the definition of the probability density in position space because there are various properties of relativistic particles that may be used to define their position. In particular, we may use the energy distribution, as we have done for photons \cite{bb}, or the rest mass distribution.

In this work we follow Dirac \cite{dir} who postulated (p.~258) that  ``The square of the modulus of the wave function, summed over the four components, should give the probability per unit volume of the electron being at a certain place''. This quantity is also (\cite{dir} p.~260) the time component $j^0$ of the electromagnetic current density $j^\mu$. The continuity equation $\partial_\mu j^\mu=0$ guarantees the conservation of the total probability,
\begin{eqnarray}\label{prob}
\int d^3r|\rho_r(\bm r,t)={\rm const},
\end{eqnarray}
where
\begin{eqnarray}\label{probr}
\rho_r(\bm r,t)=|{\bm\psi}(\bm r,t)|^2=\sum_\alpha|\psi_\alpha(\bm r,t)|^2,
\end{eqnarray}
and $\psi_\alpha(\bm r,t)$ are the components of the Dirac bispinor. Thus, the localization of the electron is determined by its interaction with the electromagnetic field: the electron is where its charge is. A natural definition of the probability density in momentum space is the modulus squared of the Fourier transform of $\psi(\bm r)$,
\begin{eqnarray}\label{probp}
\rho_p(\bm p,t)=|{\tilde{\bm\psi}}(\bm p,t)|^2=\sum_\alpha|\tilde{\psi}_\alpha(\bm p,t)|^2,
\end{eqnarray}
where
\begin{eqnarray}\label{ft}
\tilde{\psi}_\alpha(\bm p,t)=\int\frac{d^3r}{(2\pi)^{3/2}}{\psi}_\alpha(\bm r,t)e^{-\rmi{\bm p}\cdot{\bm r}/\hbar}.
\end{eqnarray}
Thus, the relation between $\rho_r(\bm r,t)$ and $\rho_p(\bm p,t)$, based on the Fourier transformation, is the same as in the nonrelativistic quantum mechanics.

From the probability densities in position and in momentum space we construct the standard expressions for the dispersions $\Delta r^2$ and $\Delta p^2$:
\begin{eqnarray}
\Delta r^2=\frac{1}{N^2}\int\!d^3r\,({\bm r}-\langle\bm r\rangle)^2\rho_r({\bm r}),\label{dr2}\\
\Delta p^2=\frac{1}{N^2}\int\!d^3p\,({\bm p}-\langle\bm p\rangle)^2\rho_p({\bm p}),\label{dp2}
\end{eqnarray}
where $N^2$ is the normalization constant,
\begin{eqnarray}\label{norm}
N^2=\int\!d^3r|{\bm\psi}(\bm r)|^2=\int\!d^3p|{\tilde{\bm\psi}}(\bm p)|^2.
\end{eqnarray}
We omitted the time dependence because the uncertainty relations are always formulated at a fixed time.

A simple textbook proof of the HUR in nonrelativistic quantum mechanics \cite{mes,bck} can be extended to the relativistic theory. This extension starts from the following non-negative expression built from the Dirac bispinor:
\begin{eqnarray}\label{old}
\int\!d^3r|({\bm r}-\langle\bm r\rangle)-\rmi\zeta(\frac{\hbar}{\rmi}{\bm\nabla}-\langle\bm p\rangle)\psi|^2\geq 0.
\end{eqnarray}
After the rearrangement, this inequality becomes a quadratic form in $\zeta$:
\begin{eqnarray}\label{pol}
Q(\zeta)=\Delta r^2+\zeta^2\Delta p^2+3\zeta\hbar\geq 0,
\end{eqnarray}
where we used the expression for $\Delta p^2$ in the position representation,
\begin{eqnarray}\label{dp21}
\Delta p^2=\frac{1}{N^2}\int\!d^3r\,|(\frac{\hbar}{\rmi}{\bm\nabla}-\langle\bm p\rangle)\psi|^2.
\end{eqnarray}
The requirement that $Q(\zeta)$ has at most one real zero (the discriminant cannot be positive) leads to the standard nonrelativistic HUR in 3 dimensions,
\begin{eqnarray}\label{hur}
\sqrt{\Delta r^2\Delta p^2}\geq\frac{3}{2}\hbar.
\end{eqnarray}
Our task is to determine whether in the relativistic theory this inequality is saturated or whether it is a strict inequality. Saturation would mean that there exists a bispinor describing an electronic state which obeys the set of three equations,
\begin{eqnarray}\label{three}
\left[({\bm r}-\langle\bm r\rangle)-\rmi\zeta(\frac{\hbar}{\rmi}{\bm\nabla}-\langle\bm p\rangle)\right]\psi=0.
\end{eqnarray}
In the nonrelativistic quantum mechanics, these equations are satisfied by a Gaussian wave function but in the relativistic case it is not so simple. Of course, a Gaussian wave function is a solution the equation (\ref{three}) but such a wave function does not describe a state of an electron but it will also have a part describing the positron. We shall analyze this problem in detail in the next Section with the use of a variational procedure.

\section{Evaluation of the variational functional}

The variational procedure is simplified if we put $\langle\bm r\rangle=0$ and $\langle\bm p\rangle=0$ in the definitions (\ref{dr2}) and (\ref{dp2}). In the case of $\Delta r^2$ the prescription is simple. We just choose the origin of the coordinate system at the center of the probability distribution. In the case of $\Delta p^2$ the solution is more subtle because the space of momenta is not homogeneous and we cannot choose its origin at will. We may, however, make the following replacement: ${\bm\psi}(\bm r)\to\exp i(\langle\bm p\rangle\cdot{\bm r}){\bm\psi}(\bm r)$ which does not restrict the freedom in the choice of the variational trial function. Such a replacement does not change $\Delta r^2$ but it eliminates $\langle\bm p\rangle$ in (\ref{dp21}), so that we obtain:
\begin{eqnarray}\label{dp22}
\Delta p^2=\frac{1}{N^2}\int\!d^3p\,{\bm p}^2\rho_p({\bm p}).
\end{eqnarray}
In this way we can eliminate completely the average values of position and momentum from the formulas for the dispersions.

We will express now the product $\Delta r^2\Delta p^2$ in terms of the independent degrees of freedom that can be subjected to unrestricted variations. The solution of the differential equation obtained from the variational procedure will determine the minimal value of $\sqrt{\Delta r^2\Delta p^2}$.

The states of an electron are described by the wave functions obeying the Dirac equation  ($c=1,\hbar=1$),
\begin{eqnarray}\label{dirac}
\left[i\gamma^\mu\partial_\mu-m\right]{\bm\psi}(\bm r,t)=0.
\end{eqnarray}
Since we are interested in the wave functions of an electron and not of a positron, we consider only positive energy solutions of the Dirac equation. Such bispinors ${\bm\psi}_\alpha(\bm r,t)$ can be expressed in terms of two complex amplitudes $f(\bm p,\pm)$:
\begin{eqnarray}\label{pw}
&{\bm\psi}_\alpha(\bm r,t)=\int\!\frac{d^3p}{(2\pi)^{3/2}}\nonumber\\
&\times\sum_{s=\pm} u_\alpha(\bm p,s)f(\bm p,s)\exp(\rmi{\bm p}\!\cdot\!{\bm r}-\rmi E_pt).
\end{eqnarray}
where $E_p=\sqrt{m^2+{\bm p}^2}$.

The wave functions $f({\bm p},s)$ of momentum variables and the spin index $s$ are not subjected to any restrictions; they represent the independent degrees of freedom of an electron moving in free space. They form a representation of the Poincar\'e group (inhomogeneous Lorentz group) for massive spin 1/2 particles, as has been described by Wigner \cite{wig}.

Two orthonormal bispinors $u_\alpha(\bm p,s)$ in the Weyl (chiral) representation of $\gamma$ matrices \cite{weyl} will be chosen in the form:
\numparts
\begin{eqnarray}\label{u}
u(\bm p,+)=\frac{1}{\sqrt{4E_p(E_p+m)}}\left[\begin{array}{c}
m+E_p+p_z\\p_x+\rmi p_y\\m+E_p-p_z\\-p_x-\rmi p_y
\end{array}\right],\\
u(\bm p,-)=\frac{1}{\sqrt{4E_p(E_p+m)}}\left[\begin{array}{c}
p_x-\rmi p_y\\m+E_p-p_z\\-p_x+\rmi p_y\\m+E_p+p_z
\end{array}\right],
\end{eqnarray}
\endnumparts
They are linear combinations of the simpler bispinors used before \cite{bb1} but they are more convenient in setting up the variational functional. The Weyl representation of $\gamma^\mu$ matrices is better suited for the relativistic analysis than the commonly used Dirac representation because the upper and lower components are then genuine relativistic spinors.

To set up the stage for the variational calculation, we shall express the dispersions in terms the amplitudes $f(\bm p,s)$. Orthonormality of the bispinors $u(\bm p,s)$ leads immediately to simple forms of and $N^2$ and $\Delta p^2$,
\begin{eqnarray}\label{dpn}
N^2&=\int\!d^3p\,\sum_s|f(\bm p,s)|^2,\\
\Delta p^2&=\frac{1}{N^2}\int\!d^3p\,{\bm p}^2\sum_s|f(\bm p,s)|^2.
\end{eqnarray}
The corresponding formula for $\Delta r^2$ is obtained with the use of the following identity:
\begin{eqnarray}\label{rel}
{\bm r}\psi_\alpha(\bm r)=\int\!\frac{d^3p}{(2\pi)^{3/2}} \rmi{\bm\nabla}_p\left(\sum_s u_\alpha(\bm p,s)f(\bm p,s)\right)\exp(\rmi{\bm p}\!\cdot\!{\bm r}),
\end{eqnarray}
\begin{eqnarray}\label{dr2c}
\fl&\quad\Delta r^2=\frac{1}{N^2}\sum_s\int\!d^3p\Big[|\nabla_p f({\bm p},s)|^2+\left(\frac{1}{p^2}-\frac{m}{p^2E_p}+\frac{m^2}{4E_p^4}\right)
|f({\bm p},s)|^2\nonumber\\
\fl&-\rmi\frac{s}{2p^2}\left(1-\frac{m}{E_p}\right)\left(p_x f^*({\bm p},s){\overleftrightarrow{\partial_y}}
f({\bm p},s)-p_y f^*({\bm p},s){\overleftrightarrow{\partial_x}}
f({\bm p},s)\right)\Big]+\frac{1}{2N^2} \int\!d^3p\frac{1}{p^2}\nonumber\\
\fl&\!\times\!\left(1-\frac{m}{E_p}\right)\left[(p_x-\rmi p_y)f^*({\bm p},+){\overleftrightarrow{\partial_z}}f({\bm p},-)-p_zf^*({\bm p},+)({\overleftrightarrow{\partial_x}}-\rmi{\overleftrightarrow{\partial_y}})f({\bm p},-)+c.c\right]\!,
\end{eqnarray}
where $\partial_x=\partial/\partial_{p_x},\,\partial_y=\partial/\partial_{p_y}\,\partial_z=\partial/\partial_{p_z}$.\\
Double arrows in (\ref{dr2c}) denote the antisymmetrized derivative, i.e. $f{\overleftrightarrow{\partial_x}}g=f\partial_x g-g\partial_x f$.
We now express these formulas in spherical coordinates to simplify the variational equations.
\begin{eqnarray}\label{na}
N^2=\int_0^\infty\!dp\,p^2
\int_0^\pi\!d\theta\sin\theta\int_0^{2\pi}\!d\phi\sum_s
|f(p,\theta,\phi,s)|^2,
\end{eqnarray}
\begin{eqnarray}\label{dp2a}
\Delta p^2=\frac{1}{N^2}\int_0^\infty\!dp\,p^2
\int_0^\pi\!d\theta\sin\theta\int_0^{2\pi}\!d\phi\sum_s
p^2|f(p,\theta,\phi,s)|^2,
\end{eqnarray}
\begin{eqnarray}\label{dr2a}
\Delta r^2=\frac{1}{N^2}\int_0^\infty\!dp
\int_0^\pi\!d\theta\sin\theta\int_0^{2\pi}\!d\phi\nonumber\\
\fl\times\sum_s\Big[p^2|\partial_pf(p,\theta,\phi,s)|^2
+|\partial_\theta f(p,\theta,\phi,s)|^2
+|\csc\theta\,\partial_\phi f(p,\theta,\phi,s)|^2\nonumber\\
\fl+\!\left(1-\frac{m}{E_p}+\frac{m^2p^2}{4E_p^4}\right)|f(p,\theta,\phi,s)|^2
-\rmi\frac{s}{2}\left(1-\frac{m}{E_p}\right)f^*(p,\theta,\phi,s)
{\overleftrightarrow{\partial_\phi}}f(p,\theta,\phi,s)\Big]\nonumber\\
+\frac{1}{2N^2}\!\int_0^\infty\!\!dp\int_0^\pi\!
d\theta\sin\theta\int_0^{2\pi}\!d\phi
\Big[\left(1-\frac{m}{E_p}\right)\nonumber\\
\fl\times\left(f^*(p,\theta,\phi,+){\overleftrightarrow{\partial_\theta}}
f(p,\theta,\phi,-)+\rmi\cot\theta f^*(p,\theta,\phi,+){\overleftrightarrow{\partial_\phi}}
f(p,\theta,\phi,-)\right)\rme^{-\rmi\phi}+c.c\Big].
\end{eqnarray}
The product $\sqrt{\Delta r^2\Delta p^2}$ of the dispersions is a well defined functional of the electron amplitudes $f(p,\theta,\phi,s)$. The minimal value of this product that appears on the right hand side of the uncertainty relation (\ref{hur}) will be found in the next Section.

\section{Solutions of the variational equations}

We will search for the function that minimizes the uncertainty relation only among spherically symmetric functions $f(p)$. In this case the variation of $\Delta r^2\Delta p^2$ with respect to $f^*$ gives the following ordinary differential equation for $f$:
\begin{eqnarray}\label{var}
\fl\Bigg[\Delta p^2\!\left(\!\!-\partial^2_p-\frac{2}{p}\partial_p+\frac{1}{p^2}
-\frac{m}{p^2\sqrt{m^2+p^2}}+\frac{m^2}{4(m^2+p^2)^2}\!\right)\!
+\Delta r^2p^2-2\gamma^2\Bigg]f(p)=0,
\end{eqnarray}
where $\gamma=\sqrt{\Delta r^2\Delta p^2}/\hbar$. We dropped the index $s$ because the equations for both components of $f(p,s)$ are identical. In terms of the dimensionless momentum variable $q$ and the dimensionless parameter $d$, the variational equation takes on the following form:
\begin{eqnarray}\label{eval}
\fl\frac{1}{2}\Bigg[-\partial^2_q-\frac{2}{q}\partial_q
+\frac{1}{q^2}
-\frac{1}{q^2\sqrt{1+d^2q^2}}+\frac{d^2}{4(1+d^2q^2)^2}
+q^2\Bigg]f_d(q)=\gamma\,f_d(q),
\end{eqnarray}
where
\begin{eqnarray}\label{dim}
q=\frac{p}{mcd},\quad
d=\frac{1}{mc}\left(\frac{\hbar^2\Delta p^2}{\Delta r^2}\right)^{1/4},
\end{eqnarray}
and we explicitly indicated the dependence of the function $f$ on the parameter $d$.

Thus, our variational equation has the form of the eigenvalue problem for the radial wave equation with a modified harmonic oscillator potential. The lowest value of $\gamma$ (the energy of the ground state) will be the best bound in the uncertainty relation. Should we allow for the angular dependence, the effective potential would pick up the centrifugal term $l(l+1)/q^2$. This would definitely increase the ground state energy.

\begin{figure}
\begin{center}
\includegraphics[width=0.8\textwidth]{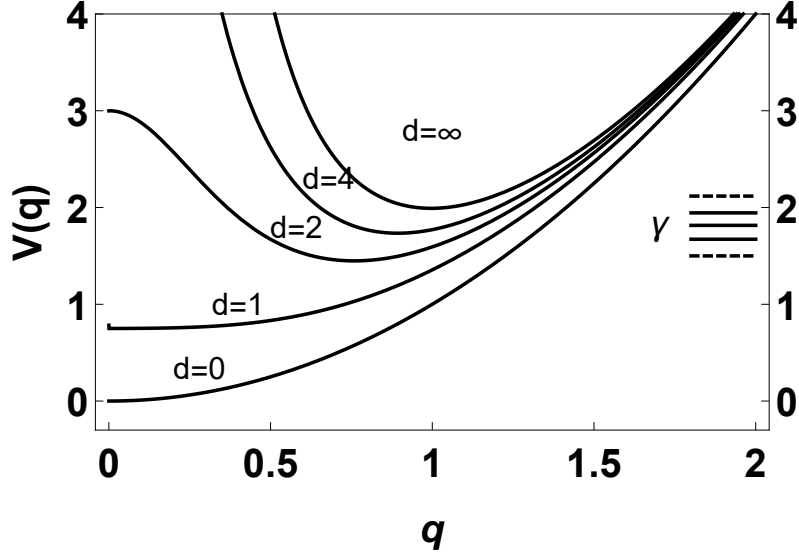}
\caption{Dependence of the potential $V(q)$ on the parameter $d$.\\
The horizontal lines mark the values of the five ground state energies $\gamma$\\
of the eigenvalue problem (\ref{eval}) obtained for the same values of $d$. The eigenvalues for the limiting cases $d=0$ and $d=\infty$ are represented by dashed lines.}
\label{fig1}
\end{center}
\end{figure}

The eigenvalue problem (\ref{eval}) has no analytic solution but considering the potential $V(q)$,
\begin{eqnarray}\label{pot}
V(q)=\frac{1}{q^2}
-\frac{1}{q^2\sqrt{1+d^2q^2}}+\frac{d^2}{4(1+d^2q^2)^2}
+q^2,
\end{eqnarray}
as a function of the parameter $d$ we find (see Fig.~\ref{fig1}) that the smaller the value of $d$, the deeper the potential. Therefore, we will obtain lower values of the eigenvalue $\gamma$ for smaller values of $d$. In the formal nonrelativistic limit (when $m\to\infty$ or $c\to\infty$) we have $d\to 0$ and the variational equation is that of the harmonic oscillator in 3D,
\begin{eqnarray}\label{eval1}
\frac{1}{2}\Bigg[-\partial^2_q-\frac{2}{q}\partial_q
+q^2\Bigg]f_0(q)=\gamma_0\,f_0(q).
\end{eqnarray}
Therefore, the wave function $f_0(q)$ is a Gaussian,
\begin{eqnarray}\label{wf1}
f_0(q)=e^{-\gamma_0 q^2/2},
\end{eqnarray}
and we obtain the nonrelativistic result $\gamma_0=3/2$.
The physical limit of $d\to 0$ is achieved when either the dispersion in $p$ tends to 0 or the dispersion in $r$ tends to infinity. Of course, these limits cannot be reached for regular, integrable functions. Thus, in relativistic quantum mechanics of electrons the HUR has the form of the {\em strict} inequality:
\begin{eqnarray}\label{hur1}
\sqrt{\Delta r^2\Delta p^2}>\frac{3}{2}\hbar.
\end{eqnarray}
In contrast to the nonrelativistic case, this inequality is never saturated. There is no square integrable function that would give the equality in (\ref{hur1}), but one may get arbitrarily close to $\frac{3}{2}$.

An exact form of the relativistic uncertainty relation can be obtained also in the other extreme case; in the ultra-relativistic limit, when $d\to\infty$. In this case, the variational equation
\begin{eqnarray}\label{eval2}
\frac{1}{2}\Bigg[-\partial^2_q-\frac{2}{q}\partial_q
+\frac{1}{q^2}+q^2\Bigg]f_\infty(q)=\gamma_\infty\,f_\infty(q)
\end{eqnarray}
again has an analytic solution. The wave function of the lowest energy state and the corresponding value of $\gamma$ are:
\begin{eqnarray}\label{wf2}
f_\infty(q)=q^{(\sqrt{5}-1)/2 } e^{-q^2/2},\quad\gamma_\infty=3/2\sqrt{1+\frac{4\sqrt{5}}{9}}=1+\sqrt{5}/2.
\end{eqnarray}
Quite unexpectedly, this value of $\gamma_{\infty}$ is the same as for photons \cite{bb}, even though the definition of $\Delta r^2$ for photons is different; it is based on the energy density and not on the charge density.

\section{Two explicit examples}

The general results presented here are well illustrated by two analytic solutions of the Dirac equation: Ground state of the electron in the Coulomb potential and the Dirac hopfion (discovered by us recently \cite{hopf}).

\subsection{Hydrogen-like ion}

The Dirac bispinor describing the ground state of the electron in a hydrogen-like ion has the form (\cite{bd} p.~55):
\begin{eqnarray}\label{bd}
\Psi_{\rm G}(r,\theta,\phi,t)=N_{\rm G}r^{\gamma-1}\rme^{-\sqrt{1-\gamma^2}r}\rme^{-\rmi Et}
\left[\begin{array}{c}
1\\
0\\
\rmi\sqrt{\frac{1-\gamma}{1+\gamma}}\cos\theta\\
-\rmi\sqrt{\frac{1-\gamma}{1+\gamma}}\rme^{\rmi\phi}\sin\theta
\end{array}\right],
\end{eqnarray}
where $\gamma=\sqrt{1-\alpha^2Z^2}$, the unit of length is the electron Compton wave length $\hbar/mc$, and the normalization constant $N_{\rm G}$ is:
\begin{eqnarray}\label{n}
N_{\rm G}=\sqrt{\frac{2^{2\gamma}(1+\gamma)^{\gamma+3/2}(1-\gamma)^{\gamma+1/2}}
{4\pi\Gamma(1+2\gamma)}}.
\end{eqnarray}
We use the standard notation for the hydrogenic wave functions, hoping that the use of $\gamma$ in this subsection will not be confused with its use in the uncertainty relations. For the bispinor (\ref{bd}) both dispersions $\Delta r^2$ and $\Delta p^2$ can be analytically evaluated and the HUR reads:
\begin{eqnarray}\label{hur2}
\sqrt{\Delta r^2\Delta p^2}=\hbar\sqrt{\frac{\gamma^2+5\gamma+2-\gamma^3}{2\gamma(2\gamma-1)}}.
\end{eqnarray}
Our parameter $d$ which characterizes the transition from the nonrelativistic to the highly relativistic regime has the form:
\begin{eqnarray}\label{pard}
d=\left(2\frac{(1+\gamma)(1-\gamma)^2(2-\gamma)}{\gamma(4\gamma^2-1)}\right)^{1/4}.
\end{eqnarray}
For hydrogen, when the Coulomb field is weak ($\gamma\approx 1,\,d\approx 0$) the r.h.s. in (\ref{hur2}) is $\hbar\sqrt{7/2}$, so it is still much higher than the lower bound $3/2\hbar$. The singularity of the Dirac wave function at the origin makes $\Delta p^2$ infinite at $\gamma=1/2$ (ultra-relativistic regime $d=\infty$). This value of $\gamma$ gives $Z=116$ which coresponds to the very end of the Mendeleev table.

\subsection{Dirac hopfion}

The localized solution the Dirac equation for a free electron $\Psi_{\rm H}$, called the Dirac hopfion \cite{hopf}, can be represented as the following Fourier integral ($\hbar=1$):
\begin{eqnarray}\label{dh}
\Psi_{\rm H}(\bm r,t)=N_{\rm H}\int d^3p\,
\tilde{\Psi}(\bm p)
e^{-\rmi E_p t+\rmi{\bm p}\cdot{\bm r}},
\end{eqnarray}
where
\begin{eqnarray}\label{dh1}
\tilde{\Psi}(\bm p)=\frac{\rme^{-aE_p}}{E_p}
\left[\begin{array}{c}
1\\
0\\
\frac{E_p-p_z}{m}\\
-\frac{p_x+\rmi p_y}{m}
\end{array}\right].
\end{eqnarray}
The integral (\ref{dh1}) can be evaluated in terms of Macdonald functions \cite{hopf} but for our purpose the Fourier representation is sufficient. The normalization coefficient can be expressed in terms of the Macdonald function $K_2$,
\begin{eqnarray}\label{norm1}
N_{\rm H}^{-2}=\int d^3p
|\tilde{\Psi}(\bm p)|^2=\frac{K_2(2ma)}{a}.
\end{eqnarray}
In the present case, we have a specific function at hand so that we cannot avoid the inclusion of the average value of the momentum $<\bm p>$ in the definition (\ref{dp2}) of dispersion.
\begin{figure}
\begin{center}
\includegraphics[width=0.8\textwidth]{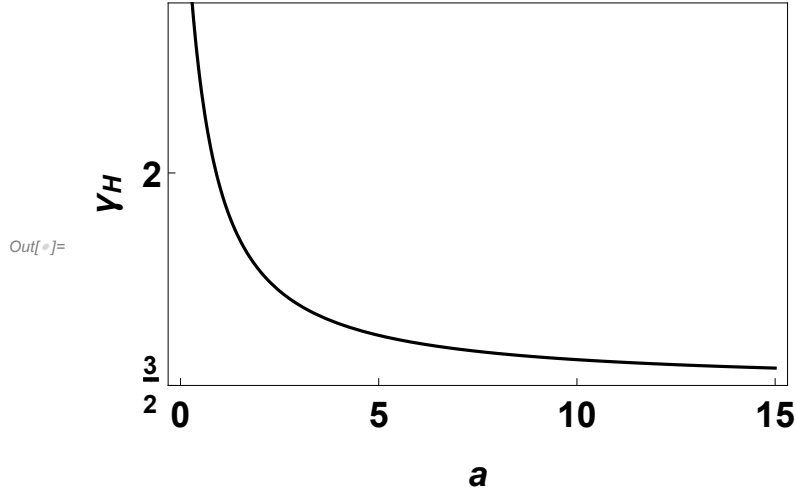}
\caption{Dependence of $\gamma_{\rm H}=\sqrt{\Delta r^2\Delta p^2}/\hbar$ on the parameter $a$.\\
The parameter $a$ is measured in units of the Compton wave length. }
\label{fig2}
\end{center}
\end{figure}
The relevant integrals in momentum space have been evaluated numerically with Mathematica \cite{math} and the results are shown in Fig.~\ref{fig2}. The nonrelativistic limit $3/2\hbar$ is approached when $a\to\infty$, i.e. when the extension of the wave packet is much larger than the electron Compton wave length.

\section{Conclusions}

We believe that we have shown that the differences between the relativistic and the nonrelativistic uncertainty relations are substantial. One may trace these differences to the presence in relativistic quantum mechanics of the scale factor, the Compton wave length $\hbar/mc$. This fact is reflected in the appearance of the dimensionless parameter $d$ in our calculations. The parameter $d$ allows one to compare directly the values of the two dispersions $\Delta r^2$ and $\Delta p^2$. One may tell that we are in the nonrelativistic regime, when the wave function is more squeezed in the momentum space than in the position space. The nonrelativistic limit is reached in two distinct physical situations, when $\Delta r^2\to \infty$ or $\Delta p^2\to 0$. The dependence of the lower bound on $d$ in the uncertainty relation, as illustrated in Fig.~1 stands in contrast to the standard nonrelativistic quantum mechanics where {\em all} Gaussians saturate the uncertainty relation.

\end{document}